# Autonomous decision-making against induced seismicity in deep fluid injections

Arnaud Mignan, Marco Broccardo, Stefan Wiemer, Domenico Giardini

The rise in the frequency of anthropogenic earthquakes due to deep fluid injections is posing serious economic, societal, and legal challenges to geo-energy and waste-disposal projects. We propose an actuarial approach to mitigate this risk, first by defining an autonomous decision-making process based on an adaptive traffic light system (ATLS) to stop risky injections, and second by quantifying a "cost of public safety" based on the probability of an injection-well being abandoned. The ATLS underlying statistical model is first confirmed to be representative of injection-induced seismicity, with examples taken from past reservoir stimulation experiments (mostly from Enhanced Geothermal Systems, EGS). Then the decision strategy is formalized: Being integrable, the model yields a closed-form ATLS solution that maps a risk-based safety standard or norm to an earthquake magnitude not to exceed during stimulation. Finally, the EGS levelized cost of electricity (LCOE) is reformulated in terms of null expectation, with the cost of abandoned injection-well implemented. We find that the price increase to mitigate the increased seismic risk in populated areas can counterbalance the heat credit. However this "public safety cost" disappears if buildings are based on earthquake-resistant designs or if a more relaxed risk safety standard or norm is chosen.

## 1. Introduction

Increasing energy needs mean increased interactions with the underground, such as fracking activities, gas extraction, waste disposal (wastewater from fracking, $CO_2$ storage), and Enhanced Geothermal Systems (EGS), all potentially inducing earthquakes (e.g., Giardini, 2009; Ellsworth, 2013; van Thienen-Visser and Breunese, 2015; Mignan et al., 2015; White and Foxall, 2016). Phasing out of nuclear energy and/or decreasing the dependence on fossil fuels also infer the increased use of alternative technologies such as EGS, or for instance $CO_2$ sequestration to cancel fossil fuel emissions. However, with increasing anthropogenic activity, larger earthquakes, damaging ones, have now become a real concern (Ellsworth, 2013; van Thienen-Visser and Breunese, 2015).

Solutions exist to limit induced seismicity, so-called traffic light systems (TLS; e.g., Bommer et al., 2006). TLSs are based on a decision variable (earthquake magnitude, peak ground velocity, *etc.*) and a threshold above which actions must be taken (e.g., stopping the injection or reducing injection rates). The definition of this threshold is so far based on expert judgment and regulations (Bosman et al., 2016). Instead, Mignan et al. (2017) proposed an actuarial approach to induced seismicity mitigation where the ATLS (for Adaptive TLS) is defined for a specific risk-based safety standard or norm. We will describe this new approach below, testing its underlying statistical model on additional EGS data. The decision-making procedure being autonomous and based on normative rules (i.e., safety standards), we can foresee a next application, which is the updating of the EGS levelized cost of electricity (LCOE) taking into account the "public safety cost", i.e., the potential cost of injection-wells being abandoned because of the ATLS. All of these steps provide the vision of an autonomous induced seismicity risk governance scheme for the geo-energy sector (Fig. 1).

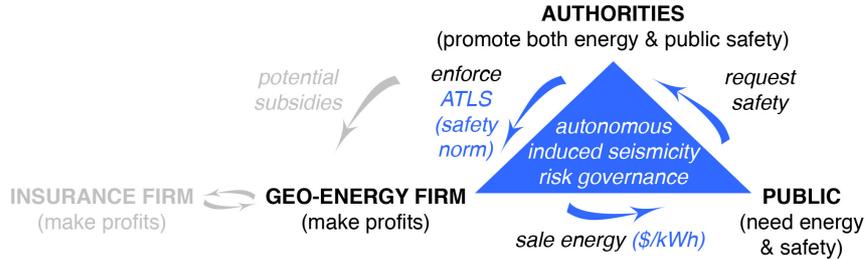

**Fig. 1:** Induced seismicity risk governance scheme for the geo-energy sector, comprising three actors: Public, authorities, and geo-energy firm. The full process runs autonomously via an ATLS and energy price updating, in blue (not considered here, in grey: potential government subsidies and insurances).

## 2. Adaptive Traffic Light System (ATLS)

### 2.1. Induced Seismicity Statistical Model

Mignan et al. (2017) presented an ATLS that evaluates the earthquake magnitude threshold $m_{th}$ not to exceed to conform with a specific risk-based safety standard or norm. This threshold can be updated in real time (Broccardo et al., 2017), based on the temporal forecasting of the induced seismicity rate $\lambda(t)$, function of the known injection flow profile $\dot{V}(t)$:

$$\lambda(t, \geq m) = \begin{cases} 10^{a_{fb}-b_{fb}m}\dot{V}(t) & ; t \leq t_{shut-in} \\ 10^{a_{fb}-b_{fb}m}\dot{V}(t_{shut-in})\exp\left(-\frac{t-t_{shut-in}}{\tau}\right) & ; t > t_{shut-in} \end{cases} \quad (1)$$

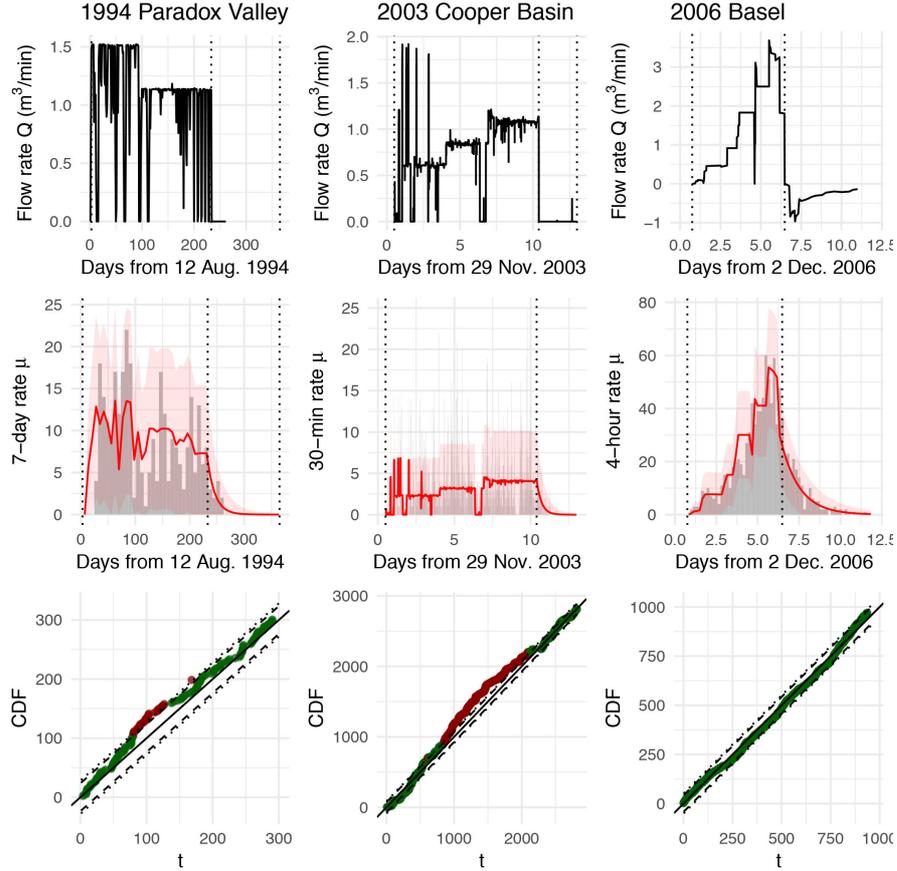

**Fig. 2:** ATLS statistical model (red curve in central row). Paradox Valley and Basel results reproduced from Mignan et al. (2017); new results for 2003 Cooper Basin dataset shown with maximum likelihood estimates $a_{fb}$ = -0.6, $b_{fb}$ = 0.97 and $\tau$ = 0.32 days. Some local variability in the data cannot be explained by Eq. (1) for constant parameter values (red dots outside the Kolmogorov-Smirnov confidence bounds on third row; see Mignan et al. (2017) for details).

The injection phase is described by a linear relationship between seismicity rate and flow rate in line with previous publications (Dinske and Shapiro, 2013; Mignan, 2016a; van der Elst et al., 2016) while the post-injection phase, after time $t_{shut-in}$, is described by a normal diffusion process (Mignan, 2015; 2016b; Mignan et al., 2017; Broccardo et al., 2017). Eq. (1) has been validated for a number of fluid injections by Mignan et al. (2017): 1994 KTB, Germany (KTB94); 1994 Paradox Valley, USA (PV94); 2006 Basel, Switzerland (B06); 2011 Garvin, USA (G11); 2012 and 2014 Newberry, USA (Nb12, Nb14). In the present study, we further test Eq. (1) on 2003 Cooper Basin, Australia (CB03; Baisch et al., 2006). Results are shown in Fig. 2 (for further statistical tests, see Broccardo et al., 2017).

It should be mentioned here that the activation feedback parameter $a_{fb}$ is analogue to the seismogenic index Σ of Dinske and Shapiro (2013). Using $a_{fb}$ is preferable to remain agnostic as to whether induced seismicity is due to poro-elasticity (Σ; Dinske and Shapiro, 2013), static overpressure (Mignan, 2016a), or any other physical process. Although $a_{fb}$ provides the average activation feedback per stimulation, a Kolmogorov-Smirnov test shows that Eq. (1) may fail locally (Fig. 2). This can be due to missing on-site information, pressure changes not linearly corelated to the injection flow rate or other second-order processes not yet considered in the model. More complex models could be implemented in the ATLS presented below, if necessary. Broccardo et al. (2017) also showed that the parameters of Eq. (1) can be updated on-line using a hierarchical Bayesian framework during stimulation (note that the framework applies also to the production phase, although the higest seismicity risk occurs during stimulation).

## 2.2. Mapping between Risk-based Safety Norm & Magnitude Threshold

Integrating Eq. (1) yields

$$\Lambda(\geq m_{saf}) = 10^{a_{fb}-b_{fb}m_{saf}}[V(t_{shut-in}) + \tau\dot{V}(t_{shut-in})] \approx Y(\geq m_{saf}) \qquad (2)$$

with $Y$ the safety standard or norm in the magnitude space, i.e. the probability of exceeding a given (relatively large) magnitude $m_{saf}$. To avoid exceeding $Y$, one must stop the injection when the following magnitude threshold is reached:

$$m_{th} = \frac{1}{b_{fb}}\log_{10}[Y - 10^{a_{fb}-b_{fb}m_{saf}}\tau\dot{V}(t_{shut-in})] + m_{saf} \qquad (3)$$

based on the condition $10^{a_{fb}-b_{fb}m_{th}}V(t_{shut-in}) = 1$, true if the injection is stopped as soon as $m_{th}$ is observed (Mignan et al., 2017).

Safety standards and norms defined for diverse hazardous environments (Jonkman et al., 2003) are often defined in terms of individual risk *IR* (i.e., the probability that a statistically representative individual dies). Such standards or norms could also be applied to geo-energy sites. For any given *IR* threshold fixed by the authorities, this value can be mapped to the magnitude space in terms of probability $Y$ of exceeding $m_{saf}$. Those parameters then depend on the risk parameters in the region of interest (mainly the seismic spatial attenuation and building type). Fig. 3 reproduces the ATLS example of Mignan et al. (2017)

where $Y(m_{saf} \geq 5.8) = 10^{-5}$ for $IR \leq 10^{-6}$ (Basel injection profile simulated with EMS-98 class A building at distance $d$ = 0 km above $z$ = 4 km borehole, and with seismic intensity attenuation relationship derived from USGS "*Did You Feel It?*" (Atkinson and Wald, 2007) corrected at the time for induced seismicity; see macroseismic risk method application to induced seismicity in Mignan et al., 2015). If the injection is stopped at $m_{th}$, the safety standard or norm is respected in average over many simulations. $m_{th}$ evolves over time since it depends on other time-varying parameters such as the $b_{fb}$ value, the ratio between small and large earthquakes, which may dramatically change during fluid injections (as was the case for the 2006 Basel experiment).

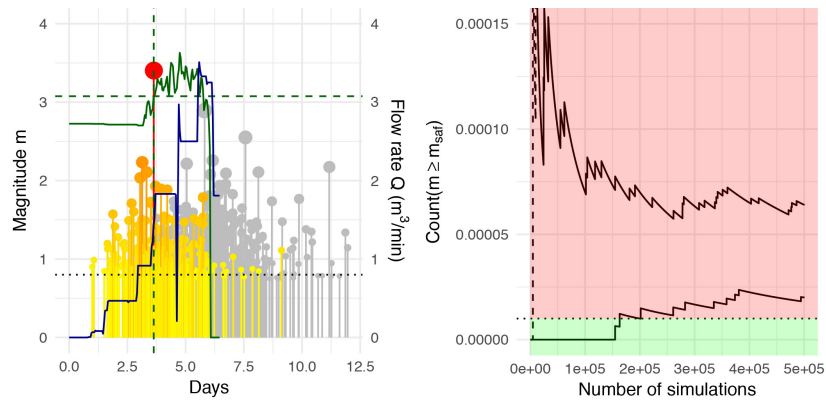

**Fig. 3:** ATLS in action. Left: Time series without ATLS (in grey) compared to a time series where the stimulation is stopped by the ATLS ($m_{th}$ in green); Right: Verification that the safety standard or norm is respected in average when the ATLS is used (bottom curve), in contrast to no ATLS (top curve) - Recomputed from Mignan et al. (2017).

The proposed ATLS is autonomous in the sense that once a safety standard or norm is selected by the authorities, the decision-making (based on Eq. 3) is done without any human intervention. All parameters and risk estimates can be estimated on-the-fly in a hierarchical Bayesian framework (Broccardo et al., 2017) and the decision ($m_{th}$) recomputed accordingly. The approach is statistically robust and transparent, which is an advantage over the standard clinical approach (e.g., Dawes et al. 1989).

## 3. LCOE Updating in ATLS Context

### 3.1. New LCOE Formulation for Different Risk Behaviours

If the ATLS were to be used systematically for a safety standard or norm selected by the authorities, one could in principle estimate the added cost of increased public safety. This is of importance in energy governance, as a trade-off must be found between public seismic safety and energy safety (produced by deep geo-energy). This can be quantified by including the "cost of public safety" in the LCOE, which corresponds to the added cost of wells abandoned due to ATLS injection termination (as represented in Fig. 3). We will assume that once a stimulation is stopped, the well will not be used again for the foreseeing future, hence requiring the use of another well. We formulate the LCOE, or price - $P = C/E$ [CHF/kWh] - as the null expectation of the following Bernoulli trial:

$$(1 - \pi)(PE - C) - \pi C_{ATLS} = 0 = \mathbb{E}[X] = (1 - \pi)x_1 + \pi x_2 \tag{4}$$

where $C$ are the standard costs [CHF] (installation, operation, and maintenance), $E$ the energy or amount of electricity produced [kWh], and $\pi$ the probability that a well will be abandoned due to the ATLS with associated costs $C_{ATLS}$ [CHF] (note that $\pi = 0$ leads back to $P = C/E$). $X = \{x_1, x_2\}$ represents the set of possible outcomes with $x_1$ representing stimulation success and $x_2$ stimulation failure due to too high seismic risk. Eq. (4) is formulated such that the outcome $x$ can be replaced by its utility $u(x)$, hence taking into account the possible risk aversion of the geo-energy firm to having a well abandoned with high *a priori* uncertainty (e.g., based on Cumulative Prospect Theory (CPT); Tversky and Kahneman, 1992). For any non-zero $\pi$, the LCOE (price $P$) increases.

**3.2. Illustrative Example**

Let us now estimate the probability $\pi$ of a well being abandoned and calculate the change in price $P$ from Eq. (4). We consider the following scenario: volume $V$ = 40,000 m$^3$ injected during planned stimulation, a maximim possible magnitude $M_{max} = 7$, and two possible safety standards or norms $IR \leq 10^{-6}$ and $IR \leq 10^{-5}$ applied for a building located at a distance $d$ from the borehole (at depth $z = 6$ km). Hazard uncertainty is defined from the underground feedback uncertainty (Table 2 of Mignan et al. (2017), plus our results for Cooper Basin - $\tau = 0$ to simplify Eq. 2) and seismic attenuation uncertainty (from Atkinson and Wald (2007) for the U.S.). Risk is then computed using the macroseismic risk approach (Lagomarsino and Giovinazzi, 2006), as applied to the induced seismicity context by Mignan et al. (2015) (incl. fatalities in Mignan et al., 2017), for two types of buildings (EMS-98 class C – reinforced concrete without earthquake-resistant design, and class D – with earthquake-resistant design). Some fatality curves are shown in Fig. 4 (left) with $\pi$ the ratio of curves failing to pass a safety standard or norm (cases for which the ATLS would stop the injection before $V$ is reached). As the seismic risk decreases with distance $d$, so does the price increase, as illustrated in Fig. 4 (right) (with $P_{base}$ = 0.35 CHF/kWh and $E$ = 1.38 10$^9$ kWh from an annual net generation of 46 GWh and a project duration of 30 years, yielding $C$ = 483 million CHF, and $C_{ATLS} = C_{well} + C_{fracturing}$ = 20.9+1 = 21.9 million CHF; taken from CH-base case of Hirschberg et al., 2015).

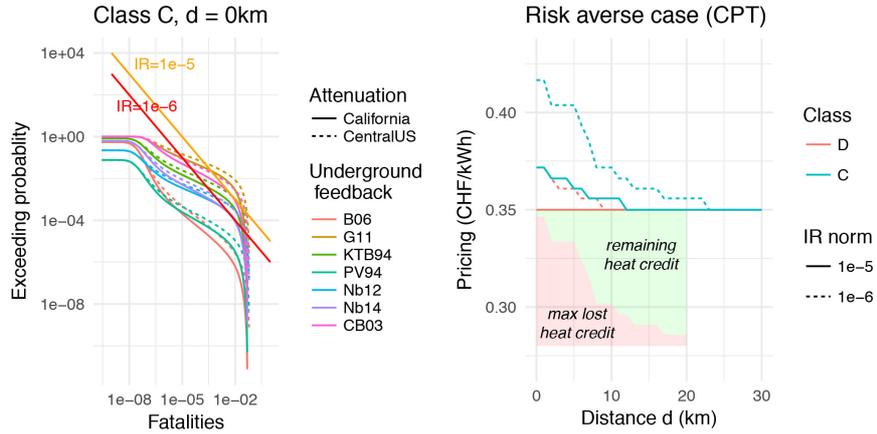

**Fig. 4:** Change in price $P$ (i.e. LCOE) due to an injection-well being abandoned with probability $\pi$ (right) computed from the ratio of fatality curves failing to pass a safety threshold (left). The price increase might counterbalance the heat credit (here 0.07 CHF/kWh; Hirschberg et al., 2015). The LCOE is further increased by risk aversion, not knowing in advance the underground feedback conditions to which the ATLS would apply. Values are illustrative only.

## 4. Conclusions

The proposed method provides the basis for an autonomous induced seismicity risk governance framework in the geo-energy sector (Fig. 1). The public requires safety from seismicity, which the authorities quantify as a risk threshold not to exceed (safety standard or norm). This can be enforced by making the geo-energy firm use the ATLS, which quantifies when a stimulation must be stopped (Eq. 3). For increased public safety comes more financial risk. This can be quantified by updating the EGS LCOE for injection-well loss (Eq. 4). The proposed actuarial/algorithmic approach could be considered in smart electricity markets (e.g., Peters et al., 2013) and in insurance smart contracts (e.g., Buterin, 2013) for induced seismicity risk.